# Challenges of measuring the impact of software: an examination of the *lme4* R package

Authors: Kai Li[1]; Pei-Ying Chen[2]; Erjia Yan[1]

[1]Drexel University, Philadelphia, PA; [2]Indiana University, Bloomington, IN

{kl696@drexel.edu; peiychen@iu.edu; erjia.yan@drexel.edu}

## Abstract:

The rise of software as a research object is mirrored in the increasing interests towards quantitative studies of scientific software. However, due to the inconsistent practice of citing software, most of the existing studies analyzing the impact of scientific software are based on identification of software name mentions in full-text publications. Despite its limitations, citation data have a much larger quantity and broader coverage of scientific fields than full-text data and thus could support findings in much larger scopes. This paper presents an analysis aiming to evaluate the extent to which citations data can be used to reconstruct the impact of software. Specifically, we identified the variety of citable objects related to the *lme4* R package and examined how the package's impact is scattered across these objects. Our results reveal a little-discussed challenge of using citation data to measure the impact of software, that even within the category of formal citation, there might be different forms in which the same software object is cited. This challenge can be mitigated by more carefully selecting objects as the proxy of software. However, it cannot be fully solved until we have one-software-one-proxy policy for software citation.

# 1 Introduction

As software "takes command" (Manovich, 2013) in every aspect of our society, so it is in the contemporary scientific practice. Upon the arrival of the era of big data, it is predicted that science is becoming increasingly reliant on statistics, as implemented in scientific software, that can analyze large amount of data. This sentiment is expressed bluntly in Chris Anderson's famous article, "The End of Theory":

> "The new availability of huge amounts of data, along with the statistical tools to crunch these numbers, offers a whole new way of understanding the world. Correlation supersedes causation, and science can advance even without coherent models, unified theories, or really any mechanistic explanation at all." (Anderson, 2008)

This new computation- and data-driven mode of scientific discovery has gradually affected nearly every scientific field. It is especially well reflected in the emerging of new concepts that combines computational methods with fields that are traditionally non- (or less-) computational, like "digital humanities" (Berry, 2011; Jones, 2013) and "computational social science" (Lazer et al., 2009; Wallach, 2016).

The rising status of software in science is a major motivation for recent scholarly interests in the scientific impact of software from a quantitative perspective. Studies have reported that software tends to be inconsistently represented in publications: authors do not cite software in consistent ways or offer enough metadata information to support the functions of software citation (Howison & Bullard, 2015; Li, Greenberg, & Lin, 2016; Li, Yan, & Feng, 2017). This reality makes it difficult to identify software from publications just based on citation data, which makes full text scientific publications a preferable data source to pursue related questions (Li & Yan, 2018; Pan, Yan, & Hua, 2016; Pan, Yan, Wang, & Hua, 2015). However, the strong reliance upon software name mentions based on full texts also limits the scope of these studies: most of these studies are conducted within full-text databases that are much more limited in the coverage of publications than citation databases.

The gap between citation data and the quantitative studies of scientific software is a major source of motivation for the present study. It is a fact that citation data does not cover the scenario in which software is mentioned but not cited in papers. However, if we assume the citation rate of software entities to be a somewhat constant variable, we can still have an informed guess of how specific software entities are used (i.e., cited and mentioned) in literature based on how they are cited, over broader scientific fields than those covered by full-text scientific publication databases. And more importantly, citation data is also accompanied by citation contexts (Small, 1982; Zhang, Ding, & Milojević, 2013), an important piece of information through which we can better understand the use of software.

However, besides its coverage issues, citations as a proxy of software in scientific publications is still under-studied. Gaps exist between citing software in scientific articles and these citations being counted as the impact of the software. In this study, we are specifically interested in one particular difficulty that has not been fully, if at all, addressed in existing studies: different objects being cited as the proxy of the same scientific software under the current software citation practice.

Two major types of citable objects related to software have been widely used. The first type is the so-called *software paper* that has been increasingly popular in scholarly communication since the early 2010s. This type of paper is often composed by software developers and offers descriptions of the software package in peer-reviewed publications. The popularity of software papers is partly driven by the fact that it helps software and software developers to be more easily credited and rewarded in the academic system through creating a paper to be cited later on (Plale, Jones, & Thain, 2014). More importantly, software papers are also designed to contribute to a better pre-publication evaluation of the software's quality and a higher degree of reusing the software package (Chue Hong, Hole, & Moore, 2013; Pradal, Varoquaux, & Langtangen, 2013). The fulfillment of these benefits is validated by the evidence that software or method papers are normally the most highly-cited ones within a database or a journal (Martín-Martín, Orduna-Malea, Ayllón, & López-Cózar, 2016; Small, 2018; Willett, 2012).

The second type of object is the software project, often in the form of a website or web page where the software is available for download. It is arguably a more natural way to represent the software than software papers and has been found to be frequently cited in scientific publications (Howison & Bullard, 2015). In the R programming ecosystem specifically, pages of individual packages on the *CRAN* or *Bioconductor* repositories are often assigned as the official citation format on the package-level (Li et al., 2017).

However, software papers and software projects are by no means the only two citable objects, and the practice to use them in publications is far from clearly defined. The study by Li and colleagues (2017) identified a highly mixed practice by researchers to cite R packages: even when a software package is instructed to be cited as a page, researchers often cite a publication instead, and *vice versa*. Moreover, it is a common practice to publish more than one software papers describing the same software package, even based on a quick examination of papers published in *the Journal of Statistical Software*, *the R Journal*, and other popular software journals.

To better understand the diversity of citable objects for the same software entity, and how this situation challenges the quantitative evaluation of the software's impact, the present study is designed to identify citable objects related to a popular software package, *lme4* in the R programming language (Bates, Mächler, Bolker, & Walker, 2015). Based on its official project page on the CRAN repository, lme4 was first deposited to CRAN in 2003[1], seven years after this repository was established (Fox & Leanage, 2016). It was designed to fit linear mixed-effects models. The following considerations contribute to the selection of this package as the research object of this study. First, it is one of the most highly-mentioned R packages based on a previous study (Li et al., 2017). Second, a number of software papers have been published focusing on this package, and its CRAN page is also a highly cited object; this offers rich data to understand how a software package is cited in different manners and disciplinary contexts over time. Last, unlike the name of "R," "lme4" could serve as a meaningful and unique query term, so that publications related to this package are more easily to be retrieved.

Based on all citable objects related to lme4, we analyzed how the impact of this package is scattered across all of its citable objects. More specifically, this study aims to answer the following two research questions.

---

[1] https://cran.r-project.org/src/contrib/Archive/lme4/

**Question 1: Which citable objects related to lme4 are cited in publications?**

This question aims to examine the variety of citable objects related to lme4. The question is directly driven by a clear gap between the official citation format that is available for every R package and the inconsistent citation practice to follow this instruction (Li et al., 2017). This inconsistency, however, cannot be easily categorized into the formal-informal citation dichotomy (Park, You, & Wolfram, 2017), because authors could be citing a CRAN page in a formal way (i.e., offering citations and references) even though what they are cited is not the official citable object.

**Question 2: How is the impact of the software package scattered across its citable objects?**

This question serves to examine the relationship among citations to different citable objects related to the same software entity, such as their overlaps and their temporal and disciplinary distributions. By conducting this analysis, we are hoping to propose a method to reconstruct the impact of lme4, at least part of it, through citation data.

This paper is organized as follows. The next section offers a review of literature related to efforts to establish standard citation formats for scientific software and quantitative examinations of the impact of software in science. It is followed by a detailed presentation of our methods and data. Our major findings are presented, with their implications discussed by the end of this article.

## 2 Literature review

### 2.1 Software citation standards

Despite the long history of software's involvement in the scientific practice (Wolfram, 1984), it is not until recently did researchers start to perceive software as a first-class research object, one that needs to be "validated, preserved, cited, and credited" (Chassanoff, Borghi, AlNoamany, & Thornton, 2018). As arguably the most important component in the contemporary academic reward system (Garfield & Merton, 1979), being citable is the prerequisite for software to hold this position.

A few efforts have been conducted to establish software as a citable object. Some of the most influential works are from the Force11 Software Citation Working Group[2]. Researchers from this group have proposed six principles that can serve as the basis for future software citation standards (Smith, Katz, & Niemeyer, 2016), including Importance, Credit and attribution, Unique identification, Persistence, Accessibility, and Specificity. These principles, of course, can be applied to both citing software as software paper and as software project, which will coexist, at least in the near future (Smith et al., 2016).

As a parallel academic genre of *data paper* (Chavan & Penev, 2011), the popularity of software papers has also been growing during recently years, which at the same time results and is resulted from the increasing number of journals that accept this type of articles (Chue Hong, 2014). The popularity of software papers can be traced to the fact that publications are the dominant currency in the current academic reward system: published articles are more likely to be cited, and thus rewarded, than other types of academic products (Smith et al., 2018). The academic publication pipeline also provides a natural solution for scientific software to be peer

---

reviewed (Chue Hong et al., 2013), similar with the argument made for the publication of datasets (Chavan & Penev, 2011).

However, one disadvantage of citing software papers is that the paper might provide citation contexts that are beyond the software per se. A software paper inevitably provides information not just about the software package, but also about its underlying theories and methods[3], and sometimes even original results (Smith et al., 2018). This makes it possible for software papers to be cited for reasons other than using the software. As shown in an earlier study (Li et al., 2016), the Plimpton-1995 paper that describes the simulation software, LAMMPS, has been frequently cited to refer to the method implemented in LAMMPS, rather than reuse of this software.

Just like dataset (Parsons & Fox, 2013), there will be no perfect metaphor to introduce software into the world of scholarly communication. Nevertheless, the present study is inspired by the disputes around the most ideal approach to software citation. We are hoping to demonstrate how the two existing approaches, namely citing software paper and citing software project, are being co-used by researchers, and how their coexistence is affecting the appreciation of the scientific impact of software.

## 2.2 Quantitative studies of scientific software

Software, as a research object, only received sporadic attentions from scientometricians before the mid-2010s. One notable earlier effort is provided by Pia and her colleagues focusing on the scientific impact of software used in high energy physics (Pia, Basaglia, Bell, & Dressendorfer, 2009, 2010, 2012). Moreover, Muenchen (2012) has offered a comprehensive overview of the popularity of major statistical software in scientific literature as well as other public venues.

However, this topic became gradually popular during the past few years. Most of these studies, notably, are based on full-text publications; they are based on either manual coding of a small sample of publications, or machine learning methods to process a larger set of data. For the first approach, Li and colleagues (2007) analyzed a sample of 400 papers mentioning the software R and examined the impact of individual R packages. More recently, Yang and colleagues (Yang et al., 2018) manually identified software entities from bioinformatic studies, and draw comparative conclusions concerning the roles played by software in bioinformatic communities in the US and China. For the second category of method, Pan and colleagues (Pan et al., 2016, 2015) developed a machine learning algorithm to extract software names from full-text scientific publications and used this method to understand the impact of software in PLoS ONE. Using the same data source, Li & Yan (2018) identified all R packages and constructed their co-mention network.

The only exception to the aforementioned approaches is a recent effort examining the impact of bibliometric mapping tools (Pan, Yan, Cui, & Hua, 2018); the authors used the Web of Science citation data to conduct their analysis.

Citation data has been proven to be an inconsistent indicator of software mentions in scientific publications (Howison & Bullard, 2015; Li et al., 2017; Pan et al., 2015). However, as a research object, its coverage is much broader than that of full-text data. The present study was proposed, in part, to demonstrate the potentials of using citation data to measure the impact of software, despite its disadvantages. Through a more careful examinations of all citable objects related to *lme4*, we are hoping to illustrate deep challenges of scientometric evaluation of software entities and some practical solutions to such challenges.

---

[3] Such as the requirement offered by the Journal of Statistical Software, that articles should be about "comprehensive open-source implementations of broad classes of statistical models and procedures or computational infrastructure upon which such implementations can be built." (https://www.jstatsoft.org/pages/view/mission)

# 3 Method and Materials

## 3.1 Selection of citable objects

We first identified all citable objects related to lme4, based on the Web of Science database and our knowledge about this package. The following three types of citable objects are identified:

- **Published articles**: articles published in journals or conference venues;

- **Unpublished articles**: articles never published in any venue; and

- **Project pages**: different web pages that host lme4.

All citable objects of lme4 are listed in Table 1, which are referred by their IDs as listed in the table during the rest of this article.

**Table 1: List of citable items of lme4**

| Type | ID | Citable item | Authors |
|------|-----|------|------|
| Published articles | PA-1 | Estimating the multilevel Rasch model: with the lme4 package | Doran, H; Bates, D; Bliese, P; Dowling, M |
| | PA-2 | The Estimation of Item Response Models with the lmer Function from the lme4 Package in R | De Boeck, P; Bakker, M; Zwitser, R; Nivard, M; Hofman, A; Tuerlinckx, F; Partchev, I |
| | PA-3 | Fitting Linear Mixed-Effects Models Using lme4 | Bates, D; Mächler, M; Bolker, B; Walker, S |
| Unpublished articles | UA-1 | Sparse Matrix Representations of Linear Mixed Models | Bates, D |
| | UA-2 | Linear mixed model implementation in lme4 | Bates, D |
| | UA-3 | Penalized least squares versus | Bates, D |

| | | generalized least squares representations of linear mixed models | |
|---|---|---|---|
| | UA-4 | Computational methods for mixed models | Bates, D |
| Project pages | PP-1 | lme4: Linear Mixed-Effects Models using 'Eigen' and S4 | Bates, D; Mächler, M; Bolker, B; Walker, S (other authors omitted) |
| | PP-2 | Mixed-effects models in R using S4 classes and methods with RcppEigen | Bolker, B; Mächler, M; Walker, S; Bates, D (other authors omitted) |
| | PP-3 | lme4 - Mixed-effects models | Mächler, M; Bates, D; Bolker, B; Walker S; Christensen, S |

The first type of object is the **published article**. All R packages have an official citation format designated by their developers. In the case of lme4, this is the paper "Fitting Linear Mixed-Effects Models Using lme4" (Bates et al., 2015) published in *the Journal of Statistical Software* (JSS). Based on the information from the JSS, this paper was submitted to the journal on June 30, 2014[4], a week after it was first deposited on ArXiv.org.[5]

In addition to this article, two other software papers have been published, both in the JSS, aiming to describe lme4:

- Estimating the multilevel Rasch model: with the lme4 package (Doran, Bates, Bliese, & Dowling, 2007)

- The Estimation of Item Response Models with the lmer Function from the lme4 Package in R (De Boeck et al., 2011)

It should be noted that all of the three papers share a similar goal, which is to demonstrate a method (i.e., Linear Mixed-Effects Models, the multilevel Rasch model, and Item Response Models, respectively) with the example of this package. The official status of PA-3 is derived from the fact that its authors are also the core developers of the package. Even though Douglas Bates is also one of the authors of PA-1, all other authors of PA-1 do not seem to be involved in the development of this package, contrary to PA-3.

---

[4] https://www.jstatsoft.org/article/view/v067i01/0
[5] https://arxiv.org/abs/1406.5823

The second category of citable object is the **unpublished manuscript**. Besides the three published articles, lme4 has also been published with a number of unpublished articles included in the package file. To identify these resources, we analyzed all versions of lme4 as deposited in the CRAN repository[6]. We examined every first minor version under each of its major versions[7]. All selected package files were downloaded and checked to identify publications included in the package. The item *PA-3* was also included in the package beginning from version 1.1-7 deposited in July 2014. Besides PA-3, four other papers have also appeared in the package, none of which has been published. The period of time in which they were included in the software package is summarized in Table 2.

**Table 2: Distribution of unpublished articles over lme4 versions**

| Major version | Date of first minor version | UA-1 | UA-2 | UA-3 | UA-4 |
|---|---|---|---|---|---|
| 0.2 | 2003/06 | | | | |
| 0.3 | 2003/07 | | | | |
| 0.4 | 2003/07 | | | | |
| 0.5 | 2004/04 | | | | |
| 0.6 | 2004/06 | X | | | |
| 0.8 | 2005/01 | X | X | | |
| 0.9 | 2005/02 | X | X | | |
| 0.95 | 2005/04 | | X | | |
| 0.96 | 2005/06 | | X | | |
| 0.98 | 2005/07 | | X | | |
| 0.995 | 2006/01 | | X | | |
| 0.9975 | 2006/10 | | X | | |
| 0.99875 | 2007/05 | | X | | |
| 0.999375 | 2008/06 | | X | X | X |
| 0.999999 | 2012/06 | | X | X | X |
| 1.0 | 2013/09 | | | X | X |
| 1.1 | 2014/03 | | | X | X |

In terms of the content, these four unpublished papers can be broadly categorized into two groups. UA-1 and UA-3 focus on the methods implemented in lme4. Specifically, UA-1 describes how to represent linear mixed-effects models using a sparse semidefinite matrix, a major functionality of lme4. UA-3, however, compares the penalized least squares approach to representation of linear mixed models, as implemented in lme4, with the generalized least squares approach. In the second group, both UA-2 and UA-4 demonstrate how to use lme4 to perform its central functions; this falls into the concept of *vignette*, i.e., detailed descriptions of the processes to conduct a certain task involving multiple functions (Gentleman et al., 2004).

The last type of citable object is the **repository page** of lme4. As mentioned above, lme4 has a page on CRAN, the official R package repository, which is titled "lme4: Linear Mixed-Effects Models using 'Eigen' and S4"[8]. Moreover, lme4 also has a project page on Github[9], titled "Mixed-effects models in R using S4 classes and methods with RcppEigen," and a page on R-Forge repository[10], titled "lme4 - Mixed-effects models." The first two pages are still active, according to their activity logs; while the R-Forge page is no long actively maintained.

### 3.2 Acquisition of citation data for citable objects

We acquired citation data for all the citable objects identified from the previous step from three sources:

- Web of Science Cited Reference Search

- Web of Science Data Citation Index (DCI)

- Google Scholar

We used both Web of Science (WoS) and Google Scholar (GS) in order to gain a more comprehensive evaluation of the impact of the software, given the facts that every database has its biases towards certain scientific fields and academic genres (Bergman, 2012; Falagas, Pitsouni, Malietzis, & Pappas, 2007; Harzing & Alakangas, 2016) and the use of software is supposedly scattered across many scientific fields (Schickore, 2017).

In Web of Science, we searched the title of each item (and in the cases of the project page, we used the page link in addition to the title) in both "Cited Work" and "Cited Title" fields in the "Cited Reference Search" tab. According to the WoS definitions, "Cited Work" refers to "cited journals, cited conferences, cited books, and cited book chapters," [11] while "Cite Title" refers to the title of the cited item per se.[12] It is assumed that both titles of the software paper and project page should be indexed as "Cited Title." However, in reality, we found that many titles, even those of published articles, can be indexed as "Cited Work" instead, which is why we used both searching methods to collect data. For all results that were retrieved, we integrated the counts of search results and all citations to them, under each citable object.

By comparison, we only searched the title of each item and recorded the citation count of the "master record" (i.e., the record that ranks the first and with the most citation counts) on Google Scholar. The master record of PP-1 is shown in Figure 1. It should be noted that this record was

---

clearly merged with other citable objects in the database, as illustrated by the PDF file format and the web page. However, we determined it to be the CRAN page because of the correct title, author information, "R package version" note, and the fact that no other record with this combination of information has received a similar amount of citations.

Moreover, it should also be noted that in both Web of Science and Google Scholar, different formats for the CRAN project page can be found. In WoS Citation Search, this is a smaller issue, given that we only considered results that are returned for the query terms. On Google Scholar, however, many parallel formats for the project pages exist (one example is shown in Figure 1); and because of the searching mechanism of Google Scholar, it is impossible to evaluate the exact number of such results. For this reason, we did not consider any of these records, even though many of them have received substantial citations.

**Figure 1: The master record (above) and a parallel format (below) for the lme4 CRAN page on Google Scholar**

We also used the DCI as a data source, which was recently released as part of the Web of Science services (Force & Robinson, 2014), in order to evaluate the coverage of citations for software packages indexed by this service. DCI has indexed all R CRAN pages as the proxy of R packages (shown in Figure 2). As such, we compared the citation count received by the lme4 CRAN page on DCI with those from other sources.

**Figure 2: Screenshot of the record of lme4 in the DCI**

Based on the results acquired from these databases (see Section 4.1), we combined citations received respectively by the two most highly-cited objects, *PA-3* and *PP-1*, and analyzed how the patterns of citations to both objects have changed over time and distributed across different scientific fields. We only selected these two objects because they cover most of the citations we collected, thus are able to show a trend that well represents how the software is cited. Moreover, because of the similar amount of citations they have received, their results are more meaningful to be compared with each other than with those of other objects.

Data collection was conducted on October 15, 2018. All analysis and visualization were done using the software R (R Core Team, 2016).

# 4 Results

## 4.1 How citations are scattered across citable objects of lme4
This section presents results concerning the distribution of citations to different citable objects related to lme4. The results, based on Web of Science and Google Scholar, are shown in Table 3. We list the numbers of citations ("Citation") and retrieved records ("Record") from both "Cited Work" and "Cited Title" search on Web of Science, as well as the total amounts from these two searches.

**Table 3: Citation count of representations related to the lme4 package based on Web of Science and Google Scholar**

| ID | WoS Cited Work | | WoS Cited Title | | WoS | | GS |
|---|---|---|---|---|---|---|---|
| | Record | Citation | Record | Citation | Record | Citation | Citation |
| PA-1 | 0 | 0 | 4 | 53 | 4 | 53 | 117 |
| PA-2 | 0 | 0 | 3 | 79 | 3 | 79 | 147 |
| PA-3 | 10 | 39 | 79 | 7,379 | 89 | 7,418 | 10,960 |
| UA-1 | 1 | 4 | 0 | 0 | 1 | 4 | 9 |
| UA-2 | 13 | 102 | 4 | 4 | 17 | 106 | 226 |
| UA-3 | 0 | 0 | 1 | 1 | 1 | 1 | 4 |
| UA-4 | 7 | 24 | 12 | 45 | 19 | 69 | 95 |
| PP-1 | 41 | 3,162 | 78 | 995 | 119 | 4,157 | 14,693 |

It can be easily observed that nearly every citable object has more than one results retrieved from WoS.[13] An important reason for the parallel records is that a wrong citation format was used in the publication, which was then unable to be integrated with the correct format. In other words, if everything is indexed consistently, one object should have just one record across both searches. But this situation is especially discernible for PP-1, which has over 100 cited forms. As shown in the example of some parallel records retrieved from the "Cited Work" search (as shown in Figure 3), these citation forms failed to be integrated because of either different author names (column 1), cited work (column 2), cited title (column 3), publication year (column 4), or a combination of these differences. Moreover, despite the seemingly mutual-exclusive definitions of "Cited Work" and "Cited Title" mentioned above, in reality, titles of many of these citable objects are indexed into either field by WoS. The same situation applies to Google Scholar, even though results related to the parallel records are impossible to be studied because of the reasons stated above.

---

[13] In the case of published articles, the citation count from the "Basic Search" page is only one data point within the "Cited Title" results.

**Figure 3: Parallel records for PP-1 from the "Cited Work" search**

| [Anonymous] | LME4 LINEAR MIXED EF | | 2007 |
|---|---|---|---|
| Bates, B.<br>＋ [Show all authors] | IME4 LINEAR MIXED EF | | 2013 |
| BATES, B.<br>＋ [Show all authors] | LME4 LINEAR MIXED EF | URL: http://cran.r-project.org/package<br>=lme4 | 2010 |

It should also be noted that PA-3, albeit an officially published work, also has a large number of variants due to the inconsistencies of information captured by WoS. For one thing, many papers cited the ArXiv version of this paper; they might mistakenly cite different ArXiv identifiers, which are indexed as different "cited works" in this database. Moreover, many citations included wrong journal names, volumes, or page numbers.

Both the variety of citable objects and the existence of parallel records contribute to the difficulties of tracking the impact of lme4. In terms of the first factor, the multiplicity of citable objects, citations to either PA-3 or PP-1 only cover about half of the total citations received by all citable objects. Moreover, this difficulty is heightened by the fact that many citable objects are indexed into parallel forms. The indexing errors identified above are not surprising in light of research focusing on the errors of citation databases (e.g., Meho & Yang, 2007; Vieira & Gomes, 2009; Winter, Zadpoor, & Dodou, 2014). However, these errors have stronger impacts on the evaluation of software citation than that of publication citation. Combining together, both factors significantly impact the effectiveness of services that are based on specific cited form(s) of software, such as the DCI database. Even though DCI indexed the CRAN page of lme4, it has only captured 259 citations to this item, about 6% of all citations received by the CRAN page based on results from Web of Science.

### 4.2 Comparisons of citations to different citable objects

This section presents results related to the second question, i.e., how citations to all lme4 citable objects are distributed over time and scientific fields. We selected two citable objects of lme4, PA-3 and PP-1, because they received most of the citations of all citable objects (see Table 3).

However, the sum of citations from the two searches on Web of Science should not be understood as the total amount of citations lme4 has received. This is because there might be citing papers indexed under multiple forms of the same citable object or those that co-cite different citable objects, both of which inflate the number of citations to lme4, if we simply add all citation counts together. We evaluated both possibilities by counting citing articles that fall into each of these two scenarios.

We first identified articles indexed multiple times under PA-3 and PP-1, respectively. In this regard, PA-3 has 35 repetitive citing papers, as compared to the 561 repetitive ones for PP-1.

Even though we cannot give a clear explanation as to why so many papers are repetitively indexed under the different forms of the same citable object, it is very likely that they have been mistakenly indexed by Web of Science.

As compared to the aforementioned scenario, both objects are co-cited for much fewer times among all citing articles. Based on WoS, we only found three papers citing both PA-3 and PP-1. This could suggest that both items are more likely to be cited alone in publications. This is supported by the fact that other citable objects of lme4 are also not frequently co-cited with either PA-3 and PP-1. For example, UA-2 has been co-cited with PA-3 in three papers and with PP-1 in four papers, as compared to the total 105 papers in which it is cited. UA-4 has a similar co-citation rate: among all 69 papers that cited UA-4, two and four papers also cited PA-3 and PP-1, respectively.

Figure 4 shows how citations to both PA-3 and PP-1 distribute over time, based on the data from Web of Science (on the left panel) and Google Scholar (on the right panel). Despite their difference in scale, the two graphs show a similar trend. Citations to the CRAN page accumulated stably until the software paper was published and assigned as the official citation format around 2014 and 2015. Since then, citations to the software paper has significantly increased, while the CRAN page has been decreasingly cited.

**Figure 4: Distribution of citations over time (left panel: citations on Web of Science; right panel: citations on Google Scholar)**

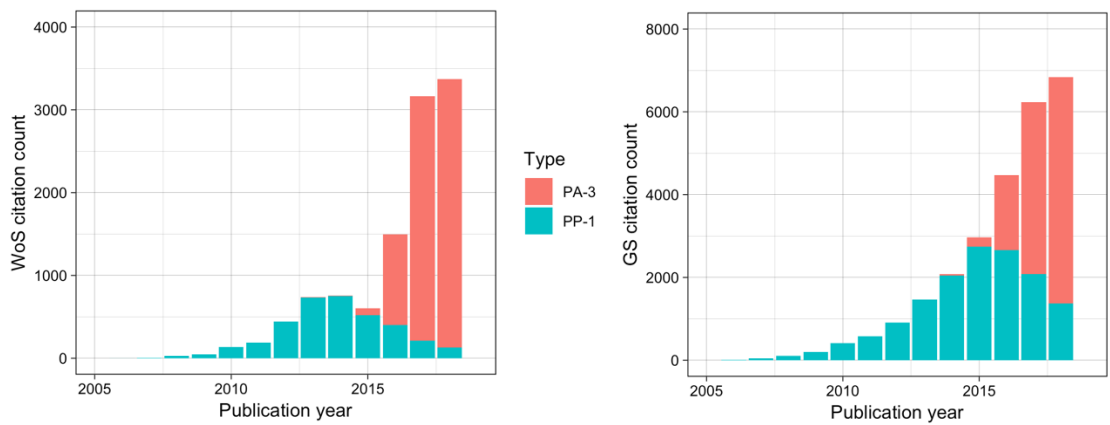

We also analyzed the top research areas in which the two items have been cited. Table 4 shows that both items have been cited in highly similar disciplinary contexts, with a concentration in biological and psychological sciences. All but two among the top 10 research areas from which they received citations are the same and with similar ratios; moreover, the top five are in the same order. For the two that are not shared by both lists, linguistics (#6 in PP-1) ranks 12th for PA-3 while neurosciences neurology (#8 in PA-3) ranks 13th for PP-1.

**Table 4: Top ten WoS areas in which PA-3 and PP-1 have been cited**

| PA-3 | | | PP-1 | | |
|---|---|---|---|---|---|
| **Area** | **Count** | **Ratio** | **Area** | **Count** | **Ratio** |

| environmental sciences ecology | 2,093 | 0.282 | environmental sciences ecology | 1,278 | 0.309 |
|---|---|---|---|---|---|
| psychology | 1,056 | 0.142 | psychology | 797 | 0.193 |
| zoology | 772 | 0.104 | zoology | 535 | 0.13 |
| science technology other topics | 754 | 0.102 | science technology other topics | 372 | 0.09 |
| evolutionary biology | 530 | 0.071 | behavioral sciences | 360 | 0.087 |
| behavioral sciences | 453 | 0.061 | linguistics | 349 | 0.084 |
| plant sciences | 403 | 0.054 | evolutionary biology | 324 | 0.078 |
| neurosciences neurology | 390 | 0.053 | biodiversity conservation | 232 | 0.056 |
| biodiversity conservation | 364 | 0.049 | life sciences biomedicine other topics | 187 | 0.045 |
| life sciences biomedicine other topics | 348 | 0.047 | plant sciences | 175 | 0.042 |

We also looked at two "abnormal" subgroups, i.e., all articles citing PA-3 published in 2015 and 2016 (*early adopters*), and those citing PP-1 published in 2017 and 2018 (*later abandoners*). The formal group is composed of 1,204 articles and the latter is composed of 508 articles. The disciplinary patterns of the two groups, again, show a strong consistency. The top five research areas of these two subgroups are displayed in Table 5. Underlying the overall consistency, however, it seems that both psychology and linguistics, the two research areas that are more strongly connected to social science, have slightly higher percentages of papers in the late abandoner group than the overall PP-1 group. This could suggest that researchers in these two research areas might be relatively slower to adopt the new citation format than those in other fields.

**Table 4: Top five WoS areas in *early adopters* of PA-3 and *later abandoners* of PP-1**

| PA-3 *early adopters* | | | PP-1 *later abandoners* | | |
|---|---|---|---|---|---|
| **Area** | **Count** | **Ratio** | **Area** | **Count** | **Ratio** |
| environmental sciences ecology | 311 | 0.258 | environmental sciences ecology | 132 | 0.26 |
| psychology | 160 | 0.133 | psychology | 111 | 0.219 |
| science technology other topics | 142 | 0.118 | zoology | 68 | 0.134 |
| zoology | 133 | 0.11 | linguistics | 52 | 0.102 |

| evolutionary biology | 94 | 0.078 | science technology other topics | 44 | 0.087 |

# 5 Discussion

## 5.1 Diversity of software citation practice and its real-world implications

In this paper, we dissected a deep challenge of using citation data to track the scientific impact of software entities, that the impact of software can be widely scattered across many different citable objects. These citable objects include not only published software papers and software project pages, but also unpublished manuscripts created along the history of the  software package. Despite the fact that we only analyzed a single software entity, lme4, the existence of multiple citable objects for the same software package is by no means a special case based on past research (Li et al., 2017). As a result, our results are applicable to many other software packages.

Our results prove that the coexistence of the practices to cite software paper and cite software project is a barrier to accurately measuring the impact of lme4. As is shown in Table 3, both PA-3 and PP-1 have received significant amounts of citations, yet, very few papers co-cited both items. It is obvious that if we only count either one of these citable objects, the impact of lme4 will be greatly underestimated. Moreover, there are other lme4-related citable objects frequently cited by researchers by a normal standard. For example, UA-2, an unpublished article, has been cited for over 100 times based WoS. Even though these objects are not pursued in this paper as detailed as the other two "major" citable objects, they must be taken into consideration in efforts to track the impact of software entities.

Besides the diversity of citable objects, another challenge of counting citations to lme4 comes from the fact that most citable objects of lme4 have multiple cited forms, which is likely derived from databases' inabilities to merge wrong citation formats. If we only search the title of the citable object in either "Title" or "Work" field in the Cited Reference search in Web of Science, again, we will miss a large chunk of citations pointing to lme4.

Moreover, these forms may have different levels of visibility in these databases, depending on which type of objects they represent. For example, a reference to a published article, even with some inaccurate information, is more easily retrievable than that to the software project page. This is exemplified by the fact that even though the lme4 page on R-Forge and Github have many mentions on Google Scholar, the records for these pages are not easily findable on either database. In other words, these two project pages have not become citable objects even on a basic level. But even the most citable project page, the CRAN page, still suffers from the infrastructure's limited capacities to handle this type of citation data. Largely because of a lack of standard citation format, the inclusion of the version number, and database's failure to integrate this flexible format, records of lme4 CRAN page are highly duplicated in both databases we examined. In Google Scholar, because of the ways in which results are presented, it is even impossible to estimate the number of records referring to this page. Both factors make services like DCI only able to capture a very small percentage of the software's impact.

Granted, such problems identified in this study can be mitigated by some practical solutions. For example, the visibility of some citation forms can definitely be improved by better integration capacities of and greater support for non-publication objects from major citation databases. However, the ultimate problem, i.e., the same software being cited as different citable objects, can only be solved when there is only one citation format for each software entity, one that represents software in a unique, persistent, and accessible way (Smith et al., 2016) being used in scientific outputs.

### 5.2 The shift of citation practice of lme4

To answer the second research question, we strived to reconstruct the impact of lme4 by locating citations to the two most frequently cited objects, PA-3 and PP-1, within a temporal and disciplinary framework. Due to the limitations of the existing citation infrastructure, this process of manually selecting citable objects is inevitable.

One of our major findings from this process is that citation patterns and behaviors seem to be greatly influenced by the changed citation instructions. PA-3 was introduced as the formal citation format in 2014 and officially published in 2015. These events have significantly changed the ways in which lme4 is cited. Since 2015, there has been an increasing trend for citations to PA-3 and an opposite one for PP-1; and this pattern is evident in both Web of Science and Google Scholar citation data. Based on an examination of the research areas of these citations, it is also found that these citations are given in similar scientific fields.

This finding is further strengthened by the fact that even during the transitional period of the citation practice of lme4 after 2015, *early adopters* of PA-3 (those citing PA-3 in 2015 and 2016) and *late abandoners* of PP-1 (and those citing PP-1 in 2017 and 2018) are still distributed across similar disciplines as compared to all lme4 citations, in spite of the small differences such as psychology and linguistics.

These findings from our analysis lead to the conclusion that publication of software papers and its being assigned as the official citation format are strong catalysts for changed citation behaviors. Given the growing numbers of software journals and software papers during the past few years (Chue Hong, 2014), it should be expected that similar trajectories has been happening to many other software packages. This shifting landscape of software citation makes it an urgent task to chart how software citation behaviors are changing, as more software papers are published and cited as the proxy of software packages. This task is foundational for both future quantitative studies on scientific software and software being truly transformed into a first-class research object. Our study offers one of the first pieces of evidence towards this research agenda. But we believe more studies are needed to fully understand how software is cited by researchers.

# 6 Conclusions

In this study, we analyzed how the lme4 R package, one of the most highly used software packages in scientific literature, has been cited as different citable objects, by using the citation data available from Web of Science and Google Scholar. Using the designed methods, we found that the impact of R has been expressed in a variety of citable objects, each with different numbers of cited forms. We found 10 citable objects related to lme4, ranging from published articles, unpublished articles, and project pages. The highly heterogeneous software citation practices, echoing what is happening in data citation (Stuart, 2017), make it extremely difficult to accurately measure the impact of lme4 and other software packages.

Given these difficulties, we manually selected citable objects related to lme4, in order to measure its scientific impact. Based on the two objects receiving the most citations, the Bates-2015 paper (PA-3) and the CRAN page (PP-1), we also evaluated the extent to which citations to lme4 changed over time and distributed over scientific disciplines after it was developed in 2003. Our results suggest that citations to the project page have been gradually replaced by citations to the software paper, after the paper became available and later assigned as the official citation format. Even during the transitional period, both articles are cited in highly similar disciplinary contexts. Given both facts, it can be safely assumed that the change of citation format for lme4 had a direct and comprehensive impact on the citation behaviors across many scientific fields, which is supposed to be the case for many other software packages, as more software developers are publishing software papers to seek more academic rewards.

This study, for the first time based on our best knowledge, offers an important piece of evidence towards the citations to multiple citable objects related to the same software entity, one important phenomenon that fails to be discussed in the academic literature of software citation. This is an important perspective towards the quantitative evaluation of software in science. However, two other topics should be thoroughly analyzed before we can reach this ultimate goal. First, we need to combine evidence from citation analysis with those from text-mining analyses. Citations to software are only one aspect of software's representations in scientific texts. Second, we need to better understand the contexts of citations to software papers. As mentioned above, not all citations to this type of papers suggests the use of the software, given the richness of semantics in this type of academic genre.

## Acknowledgement

This project was made possible in part by the Institute of Museum and Library Services for the project titled "Building an entity-based research framework to enhance digital services on knowledge discovery and delivery" (Grant Award Number: RE-07-15-0060-15) and "IDEASc: Integrated doctoral education with application to scholarly communication" (Grant Award Number: RE-02-14-0023-14).